\documentclass[structabstract]{aa}  
\usepackage{graphicx}
\usepackage{txfonts}
\begin{document}
\def\etal{{\rm et all. }}
\def\mpc{{\  h^{-1} \rm Mpc}}
\def\kpc{{\ h^{-1} \ \rm kpc}}
\def\kms{{\rm km \ s^{-1}}}

%
   \title{Galaxy interactions II: High density environments}

   \subtitle{}

   \author{Sol Alonso\inst{1,2}
          \and
          Valeria Mesa\inst{1,2}
   		  \and
          Nelson Padilla\inst{3}
          \and
           Diego G. Lambas\inst{1,4}
           }

   \institute{Consejo Nacional de Investigaciones Cient\'{\i}ficas y
T\'ecnicas, Argentina\\
              \email{salonso@icate-conicet.gob.ar}
         \and
ICATE, Universidad Nacional de San Juan, Argentina\\
         \and
Pontificia Universidad Cat\'olica, Departamanto de Astronom\'ia y Astrof\'isica \& Centro de Astroingenier\'ia, Chile           
         \and
IATE, OAC, Universidad Nacional de C\'ordoba, Laprida 854, X5000BGR, C\'ordoba, Argentina
             }
             
   \date{Received xxx; accepted xxx}

  \abstract
   {}
    {With the aim to assess the role of dense environments in galaxy interactions, 
we present an analysis of close galaxy pairs in groups and clusters, obtained from the Sloan Digital Sky Survey Data Release 7 (SDSS-DR7).}
    {We identified pairs that reside  
in groups by cross-correlating the total galaxy pair catalog with the SDSS-DR7 group catalog from Zapata et al. (2009). 
We classify pair galaxies according to the intensity of interaction as evidenced from the morphological appearance of the optical images.
We analyzed the effect of high density environments on different classes of galaxy-galaxy interactions
and we have also studied the impact of the group global environment on pair galaxies.}
    {We find that galaxy pairs are more concentrated towards the group centers with respect to 
the other group galaxy members, and disturbed pairs show a preference to contain the 
brightest galaxy in the groups.
The color-magnitude relation exhibits significant differences between pair galaxies and the control sample (constructed with galaxy group members without close companions), consisting in color tails with a clear excess of extremely blue and red galaxies for merging systems.
In addition, pair galaxies show a significant excess of young stellar populations with respect to galaxies in the control sample; this finding suggests that, in dense environments, 
strong interactions produce an important effect in modifying galaxy properties.
We find that the fraction of star forming galaxies decreases toward the group center; however, galaxy pairs show a more efficient star formation activity than galaxies without a close companion.

We have also found that pair galaxies prefer groups with low density global environments with respect to galaxies of the corresponding control sample. 

Blue, young stellar population galaxies prefer groups within low density global environments.
We find that this behavior is mainly driven by galaxy-galaxy interactions, which provide the fundamental physical mechanisms that drive this process.}
 
   {}

   \maketitle
%

\section{Introduction}

Galaxy properties (morphology, color, star formation, gas content, etc.) have a strong dependence on the local environment where galaxies reside. Several works found evidence that galaxies in high density local environment such as groups and clusters, show different properties than their isolated counterparts (e.g. Dressler 1980; Balogh et al. 2004; Baldry et al. 2006; 
Skibba \& Sheth 2009).  
Groups and clusters are dominated by spheroidal and gas poor galaxies, because when spiral galaxies traverse the dense local environments of clusters, stripping removes their interstellar gas making them loose their ability to form new stars (e.g Gunn \& Gott 1972; Dressler 1980).

In addition, recent observational and theoretical works support the idea that galaxy properties also depend on the global environment.
 
Regarding this topic, some authors (e.g. Balogh et al. 2004; Ceccarelli et al. 2008; Park \& Choi 2009, Padilla et al. 2010) showed that the star formation rate and colors depend on the large-scale structure.  
Using a semi-analytic model of galaxy formation, Gonzalez \& Padilla (2009) revealed important global effects;
the fraction of red galaxies diminishes in equal local density environments, when farther away from clusters and closer to voids. 
More recently, Cooper et al. (2010), helped improve the picture of these environmental dependences of galaxy properties by 
relating it to the assembly history of the galaxy hosts.  They found that for galaxies of equal age and mass, those
characterized by younger ages (later assembly, and bluer colors) were preferentially located in low density environments. 
Lacerna \& Padilla (2011) explained this result showing that 
the young stellar populations in galaxies in low density environments are a result of the uninterrupted growth of
their host halo, which is not the case for galaxies of equal mass in higher density environments, where the infall of
material can be diverted to larger neighbors.

On the other hand, observations show that mergers and galaxy interactions are powerful mechanisms 
that induce star formation (Kennicutt 1998), that may affect several properties of 
galaxies and their morphology.
These effects are strongly dependent on the local environment and epoch in which galaxy interactions occur.
Cosmological N-body simulations show that galaxies in close pairs are 
preferentially located in group environments (Barton et al. 2007). In the same direction, McIntosh et al. (2008)
found that massive mergers are more likely to occur in large galaxy groups than in massive clusters.
Heiderman et al. (2009) found a low merger rate that only contributes a small fraction of the total 
star formation rate (SFR) density of the A901/902 superclusters.
The predicted merger rates have uncertainties grow at the lowest masses and high redshifts (Hopkins et al. 2009; 2010),
and different assumptions in the modeling of galaxy mergers can also result in significant
differences in the timings of mergers, with consequences for the formation and evolution of galaxies (De Lucia et al. 2010).
Padilla et al. (2011) inferred the number of mergers during the evolution of early-type galaxies 
from $z = 1$ to the present-day, finding a descendants at $z = 0$ of lower number density than their progenitors, 
implying the need for mergers to decrease their number density by today.

Several observational previous works have analyzed the role of the local density environment on galaxy interactions. 
Lambas et al. (2003) showed that pair galaxies (with projected distance, $r_p <$ 25 kpc $ h ^ {-1}$, and relative radial velocity, $ \Delta V <$ 100 km $ s ^ {-1}$) in the field have a higher star formation activity than isolated galaxies in the same environment, with similar luminosity and redshift distributions.
Alonso et al. (2004), performed a study of galaxy pairs in high-density regions corresponding to groups and clusters. The results of this study indicate that galaxy pairs in groups are systematically redder and 
have less star formation activity than other galaxy group members with 
no nearby companions, except for pairs with separations of $r_p <15$ kpc $h^{-1}$, which show a 
significantly higher activity of star formation. 
Alonso et al. (2006) obtained two galaxy pair catalogs from the 2-degree field Galaxy Redshift Survey 
(2dfGRS, Colless et al. 2001) and from the second data release of the Sloan Digital Sky Survey 
(SDSS-DR2, Abazajian et al. 2004), finding that the star formation birth rate parameter is a 
strong function of the local environment, $r_p$ and $\Delta V$.
 Robaina et al. (2009) analyzed the enhancement in SFR as a function of projected galaxy separation 
in major mergers between massive galaxies. 
They find, on average, a mild enhancement 
in the SFR in pairs separated by projected distances $r_p <40$ kpc $h^{-1}$.
Their work shows that galaxy interactions are more effective at triggering important star formation
activity in low and moderate density environments with respect to the control sample of
galaxies without a close companion. 
Although close pairs have a larger fraction of actively star-forming galaxies and blue colors, there is a fraction 
of galaxy interactions that exhibit red colors with respect to those systems without a close companion.

Recently, Perez et al. (2009) found that in low and high local density environments, the colors and shapes of pairs of nearby galaxies are very similar to those of isolated galaxies; however, 
at intermediate densities they found significant differences, indicating that close pairs 
may have experienced a more rapid transition to the red sequence than isolated galaxies. 
Also, they suggest that at intermediate-density environments galaxies are pre-processed efficiently 
by close encounters and mergers before entering groups and clusters.
Using spectroscopic data from the Prism Multi-Object Survey (PRIMUS), Wong et al. (2011) studied isolated close galaxy pairs and compared these to a control sample 
in the redshift range $0.25 < z < 0.75$.
They found that pair galaxies have bluer colors on average than isolated galaxies with similar 
redshift and r-band luminosity distributions, suggesting an enhancement in SFR.

More recently, Ellison et al. (2010) used a sample of close galaxy pairs selected from the Sloan Digital 
Sky Survey Data Release 4 (SDSS-DR4) to investigate the effect of the environment on interactions. 
They found that in high density environments, mergers still occur but mainly without star 
formation; they also found that the bulge fraction is an increasing function 
of the nearest neighbor separation out to
at least 400 kpc $h^{-1}$. 
Previous works (Alonso et al. 2007; Michel-Dansac et al. 2008; Lambas et al. 2011 in preparation, hereafter L11) have also shown the importance of analyzing different classes of galaxy pair interactions. 
These studies show that merging pairs with strong signs of interaction show different properties (star formation rate, metallicity, color, etc.) with respect to pairs that show no evidence of distorted morphologies.

Motivated by these findings, in this paper we analyze the effect of high density environments on 
different classes of galaxy-galaxy interactions and the role played by the global environment
of the group on the properties of its number pairs.   We will study galaxy interactions within groups and clusters with the aim of assessing whether they 
play a significant role in modifying the properties of galaxies in pair systems, such as 
luminosities, colors, star formation rates, etc.
For this purpose, using data from the seventh release of the Sloan Digital Sky Survey (SDSS-DR7), we will obtain a large sample of pair galaxies within groups and clusters (from Zapata el al. 2009) and we will classify the interacting pairs according to the intensity of interaction as seen in the morphological appearance of the optical images (Alonso et al. 2007, L11).
We will study the spatial distribution of pair galaxy colors, stellar population and color-magnitude 
relations for different classes of galaxy-galaxy interactions within groups/clusters and compare them to those of group members without close galaxy companions, with the aim to provide further clues on the interplay between the high density environment and merger properties. 
Finally, we will analyze the effect of the group global environment, obtained from the distance to neighboring groups to try and understand the influence of the large-scale structure on the galaxy formation processes.

This paper is structured as follows. Section 2 describes the procedure used to construct the catalog of group galaxy pairs from SDSS-DR7. 
Section 3 analyses the luminosity galaxy ranking and the pair distribution in groups/clusters.
In Section 4 we discuss the distribution of galaxy pairs in groups and we study the colors and stellar population with respect to both group central distance and host group luminosity.
In section 5, we analyze the effect of the group global environment, and in Section 6 we summarize our main conclusions.


\section{Construction of a group galaxy pair catalog from SDSS-DR7}

  
The Sloan Digital Sky Survey (SDSS, York et al. 2000) is a photometric and spectroscopic
survey that covers approximately one-quarter of the celestial sphere.
The imaging portion of the seventh release of the SDSS (DR7, Abazajian et al. 2009) 
comprises 11663 square degrees of sky imaged in five wave-bands
($u$, $g$, $r$, $i$ and $z$) containing photometric parameters of 357 million objects. 
The main galaxy sample is essentially a smaller, magnitude limited spectroscopic sample 
(Petrosian magnitude) with 929555 galaxies of \textit{$r_{lim}$}$ < 17.77$. These galaxies span a
redshift range $0 < z < 0.25$ with a median redshift of 0.1 (Strauss \etal 2002).

We have calculated K-corrections using the publicly available 
code described in Blanton \& Roweis (2007) as a calibration for our K-corrected magnitudes.
In our previous work (Alonso et al. 2007), we performed a suitable photometric 
analysis to assess the reliability of the SDSS magnitude deblending procedure 
in merging pairs, finding negligible systematic effects.
As a spectral indicator of the mean stellar population age, 
we will use the strength of the 4000 $\AA$ break, $D_n(4000)$, defined as the ratio of the 
average flux densities in the narrow continuum bands (3850-3950 $\r{A}$ and 4000-4100 $\r{A}$, 
Balogh et al. 1999). This spectral discontinuity arises by an
accumulation of spectral lines in a narrow region of the spectrum, an effect that is important in the spectra of old stars.
In the following analysis, we also use the star formation rate normalized to the total mass in stars, $SFR/M^*$, taken from Brinchmann et al. (2004).

In our previous work (L11) on the SDSS-DR7, 
we built a Galaxy Pair Catalog (GPC) selecting pairs with relative 
projected separations, $r_{p}< 25 \rm \,kpc \,h^{-1}$,
relative radial velocities, $\Delta V< 350 \rm \,km \,s^{-1}$ and redshifts $z<0.1$.
In previous works by the team, we found that these limits are adequate to define galaxy pairs with enhanced 
star formation activity (Lambas et al. 2003, Alonso et al. 2006). 
We excluded AGNs and we also removed false identifications (mostly parts of the same galaxy 
and objects with large magnitude uncertainties). 
With these restrictions, the pair catalog in the SDSS-DR7 comprises 1959 reliable 
close galaxy pairs with $z<0.1$.
In the galaxy pair catalog obtained from the SDSS, the effects of incompleteness or 
aperture do not
introduce important biases (e.g see also Balogh et al.2004; Alonso et al. 2006).

In order to analyze in detail the properties of galaxy
interactions in high density environments, we identified the
pairs that live in groups by cross-correlating
the total galaxy pair catalog with the SDSS-DR7 group catalog
constructed by Zapata et al. (2009).
These authors identified groups using a friends-of-friends algorithm with
the same parameters corresponding to the values found by Merchan $\&$ Zandivarez (2005)
to produce a reasonably complete sample (95$\%$) with low contamination ($<$ 8$\%$).

The identification procedure consists of an improved
version of the Huchra \& Geller (1982) friends-of-friends algorithm, with
variable linking lengths of $D_{12}=D_0 R$ and $V_{12}=V_0 R$, in the direction perpendicular
and parallel to the line-of-sight, respectively, where
$D_0=0.24$h$^{-1}$Mpc and $V_0=450$kms$^{-1}$. $R$ is a spatial scaling that takes into
account the variation of the space density of galaxies in a flux limited catalog, and
is calculated from the ratio of the density of galaxies brighter than the minimum
required to enter the catalog at the mean distance of the galaxies being linked, and a
characteristic survey depth (see Merchan \& Zandivarez, 2005, for full details of the
group catalog construction method).  

In this work we will use the group luminosity as a proxy for the group mass, since
it has been shown that the virial mass is prone to large uncertainties below
$M_{vir}\simeq 3\times10^{13}$h$^{-1}M_{\odot}$ (e.g., Padilla et al., 2004).  Our
estimator is also a proxy for the group total luminosity since our group luminosity
consists of the addition of the luminosities of the four brightest galaxy members.
The other important group property for this work is the group center, which we calculate
by making a luminosity weighted average of the positions of the member galaxies.
The group catalog has 122,962 galaxies within 12,630 groups with a minimum number of 4 members
 and $z<0.1$.       

This catalog is constructed in a similar way to those of Berlind et al. (2006) and
Yang et al. (2007), and is composed of groups of similar quality as is evident from comparing
the completeness and contamination from these works and in Merchan \& Zandivarez (2005), while
at the same time it gives us the added advantage of allowing to perform the measurements of the group
properties that we need.

\begin{figure}
\includegraphics[width=60mm,height=60mm ]{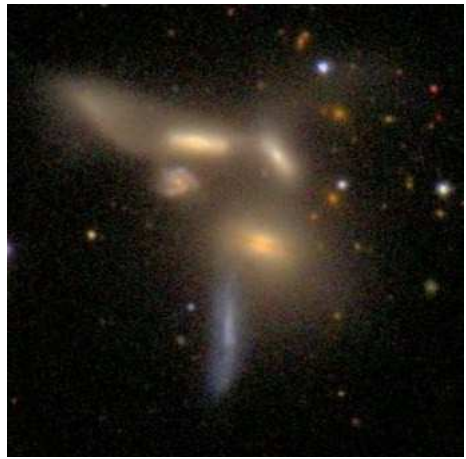}
\caption{Image of an extraordinary multiple interaction within a SDSS galaxy group.
The host group has 27 members with $M_{Virial}= 2.12*10^{13} M_{\sun}$. 
}
\label{Fusion}
\end{figure}

As a result of this cross-correlation, we obtain a sample of 660 galaxy pairs in 615 groups, with $z<0.1$. We find 40 
groups with 2 galaxy pairs, 3 groups with 3 galaxy pairs and 2 groups with 4 galaxy pairs.
Fig.\ref{Fusion} shows an extraordinary and unusual example of a multiple galaxy interaction between five group members in the local Universe. 
The host galaxy group has 27 members with $M_{Virial}= 2.12*10^{13} M_{\sun}$.

Pair galaxies in the L11 Catalog are classified as $Disturbed$ and 
$Non\ disturbed$ pairs.
The first class is further sub-classified as merging ($M$), corresponding to
pairs with evidence of an ongoing merging process,
and tidal ($T$), pairs with signs of tidal interactions but not necessarily merging.
$Non\ disturbed$ ($N$) pairs show no evidence of distorted morphologies.
For the group pair catalog we calculate the percentage of these three categories defined above
(see Table 1) and find that about 9 $\%$ of galaxy pairs are classified as 
$M$, 33 $\%$ as $T$ and 58 $\%$ as $N$.

\begin{table}
\center
\caption{Percentages of pairs classified as merging, tidal and non disturbed,
in SDSS groups}
\begin{tabular}{|c c c| }
\hline
Classification & Number of pairs &  Percentages  \\
\hline
\hline
Merging              &  58      &    8.8$\%$     \\
Tidal                &  221     &   33.5$\%$     \\
Non Disturbed        &  381     &   57.6$\%$     \\
\hline
\end{tabular}
{\small}
\end{table}

Our samples can include spurious pairs.
Mamon (1986; 1987) found that spurious pairs are expected to be more frequent in high density regions. 
Our choice of limiting values for the projected separation and radial velocity 
help to reduce this problem.
We expect the $M$ and $T$ pairs to be mostly free of contamination due to their distorted morphologies, 
but the $N$ pairs are prone to this 
problem.
In order to estimate the fraction of spurious pairs in our
galaxy groups, we ran several Monte-Carlo simulations of groups/clusters
with more than 20 galaxies in $r_{\rm p} <1 h^{-1}$ Mpc.
Our test shows that 98.2$\%$ of $M$ and $T$ pairs 
and 86.5$\%$ of $N$ pairs are real systems.
From these estimates, we conclude that although spurious pairs
are present in $N$ systems within dense regions, 
actual binary galaxies clearly dominate the statistics. 

\begin{figure}
\includegraphics[width=70mm,height=90mm ]{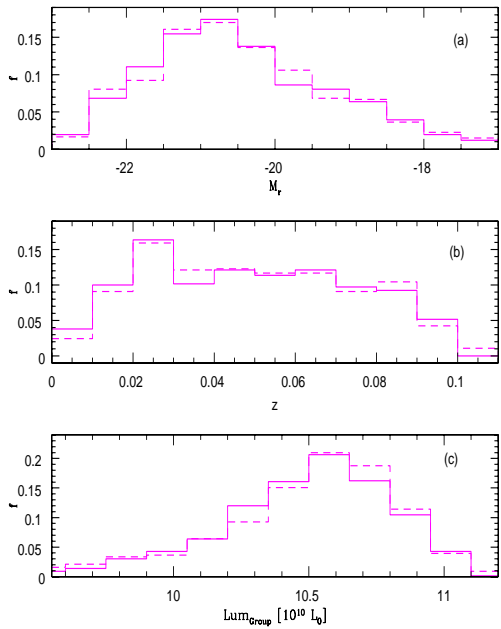}
\caption{Upper and medium panels ($a$ and $b$) show the distributions of $M_r$ and z in pair galaxies (solid lines) and in the control sample (dashed lines).
The lower panel ($c$) shows the group luminosity distribution for pair and control galaxies 
(solid and dashed lines, respectively). }
\label{histcont}
\end{figure}

We focus our attention on the effects of interactions in groups by comparing
with a suitable control sample which differs from the group pair catalog only in 
the fact that galaxies in groups in the control sample have no close companions.
Using a Monte Carlo algorithm for each group galaxy pair, we select two members of the group catalog.
Therefore, in this paper, the control sample
contains 1320 galaxies in groups and clusters which
do not have a companion within  $r_{\rm p} <100 h^{-1}$  kpc and
$\Delta V <  350 {\rm km/s}$. 
This comparison sample of galaxies in groups shares 
similar environment conditions, range of redshift, $M_r$ and group luminosity 
distributions as the sample of galaxy pairs in groups (see Fig. \ref{histcont}).

\section{Luminosity galaxy ranking and pair distribution in groups}

The core of a cluster is a hostile region with an extreme environment that favors the 
formation of less common galaxy types (Zwicky \& Humason, M. L., 1960).
The best evidence of merger processes within clusters is the presence of a giant cD galaxy in the centers of clusters (Morgan et al. 1975) which have extraordinary luminosities and masses, well above those of ordinary galaxies. The theory commonly accepted is that cD galaxies have cannibalized neighboring galaxies, accreting companions that stray into their domain. In addition, both observation and theory suggest that this process was more efficient in the past, when clusters were formed from smaller groups of galaxies with lower relative velocities.

 More recently, von der Linden et al. (2007) found that the brightest group and cluster galaxies (BCGs) are larger and have higher velocity dispersions than non-BCGs of the same stellar mass, suggesting that BCGs contain a larger fraction of dark matter.
In addition, the peculiar velocity of the BCG is independent of cluster richness and only slightly dependent on the Bautz-Morgan type (Coziol et al. 2009).
Zibetti et al. (2005) found that the surface brightness of the intracluster light (ICL) correlates 
 both with the BCG and the cluster richness, while the fraction of total light in the ICL is almost independent of these quantities. Also, the ICL is a ubiquitous feature for clusters with a single, dominant BCG, demonstrating 
 that it dominates the combined luminosity of the BCG+ICL components (Gonzalez et al. 2005).
 More recently, Skibba et al. (2011) have shown that frequently ($25\%$- $40\%$), the brightest galaxies in groups or clusters are not the central galaxies of the systems.

\begin{figure}
\includegraphics[width=90mm,height=90mm ]{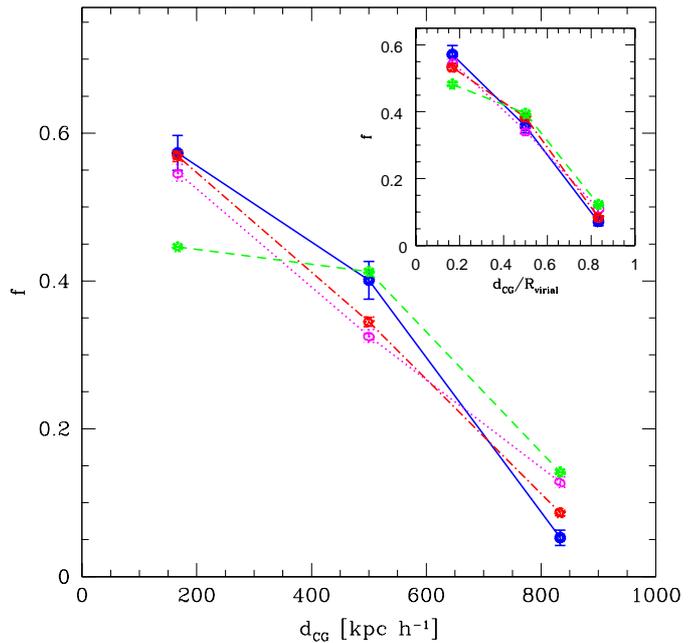}
\caption{ Distribution of group-centric galaxy distances, $d_{CG}$, for
$M$ (solid lines), $T$ (dotted lines) and $N$ (dot-dashed lines) pairs.
The dashed line represents all the group galaxy members without near companions.
The small box corresponds to the distribution of the normalized group-centric distance 
$d_{CG}/R_{Virial}$ in the same samples.
We have considered groups with a minimum number of 10 members.
}
\label{histdc}
\end{figure}

In this section, we analyze the distribution of projected radial 
distance for pairs and for all group galaxy members without near companions 
with respect to group centers.
In all samples, we have considered groups with a minimum number of 10 members. 
The resulting distributions are shown in Fig.\ref{histdc}.
It can be appreciated that galaxy pairs ($M$, $T$ and $N$) are more concentrated towards the group center with respect to the other group members.
We have also calculated the normalized group-centric distance ,$d_{CG}/R_{Virial}$, finding same 
tendencies (see the inset).
Similar results were found by Ellison et al. (2010), who analyzed the dependence of interactions on cluster-centric 
distance, finding that for close pairs the fraction of asymmetric galaxies (associated with tidal disruptions) is highest in the cluster centers. 
Moreover, Lin et al. (2010) show that the galaxy pair fraction increases with density.

\begin{figure}
\includegraphics[width=80mm,height=80mm ]{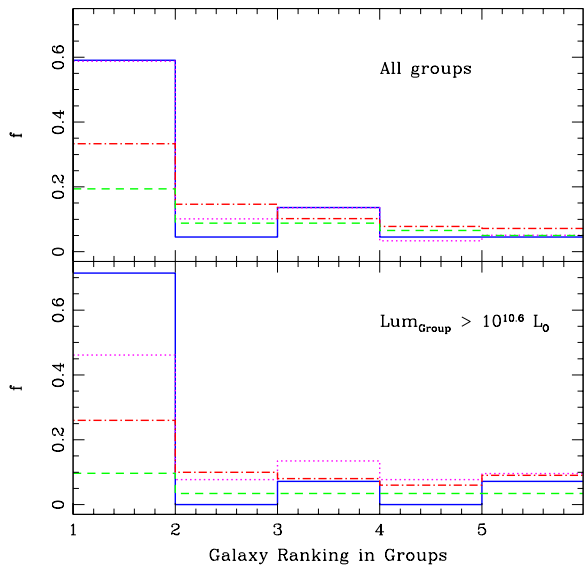}
\caption{ Galaxy $r-$band luminosity ranking for different pairs classified as $M$, $T$ and $N$ 
(solid, dotted and dot-dashed lines, respectively) in all groups, with a minimum number of 4 members
(upper panel) and in groups with $Lum_{Group} > 10^{10.6} L_{\sun}$ (lower panel).
The dashed lines represent all the group galaxy members without near companions.
The number 1 indicates the most luminous galaxy in the group, 
number 2 the second most luminous galaxy, and so on.}
\label{HRank}
\end{figure}

We also study the luminosity galaxy ranking in pairs within groups (Fig.\ref{HRank}).
The luminous galaxy in the group is ranked 1$^{th}$, and so on for the   second, third, etc, most luminous galaxies.
It can be appreciated that disturbed
pairs ($M$ and $T$) have a preference to contain the brightest group galaxy 
(45$\%$ and 38$\%$ respectively, in all groups;  
63$\%$ and 32$\%$ in groups with $Lum_{Group} > 10^{10.6} L_{\sun}$) with respect to non disturbed pairs and other group galaxy members without near neighbors (see Table 2). 
These behaviors indicate that morphologically disturbed galaxy pairs are 
more concentrated in the central regions of the groups/clusters and 
contain preferentially the brightest object in the group.
These results are in agreement with the current paradigm of galaxy formation which
assumes that the structures form hierarchically, such that smaller halos merge to form larger and more massive halos via, mostly, minor mergers.
Consistent with the latter, the central galaxy is generally the most luminous, most massive galaxy in a dark matter halo and it resides at the center of the potential well of the halo; and is also expected to cannibalize neighbor satellites.

\begin{table}
\center
\caption{Probability of different galaxy pair classes to be the brightest galaxy 
in the group/cluster}
\begin{tabular}{|c c c| }
\hline
Pair           & $\%$ in all groups  & $\%$ in groups with   \\
Classification &                     & $Lum_{Group} > 10^{10.6} L_{\sun}$ \\
\hline
\hline
  $M$      &  45$\%$      &    63$\%$     \\
  $T$      &  38$\%$      &    32$\%$     \\
  $N$      &  24$\%$      &    14$\%$     \\
  $GGM$    &  10$\%$      &    5$\%$     \\
\hline
\end{tabular}

{\small Note: $GGM$: group galaxy members without near companions}
\end{table}

\section{Colors and stellar population}

Different authors (Baldry et al. 2004, Balogh et al. 2004) showed that the  
distribution of galaxy colors could be well fit by a double Gaussian with  
blue and red peaks over a wide range of absolute magnitudes and local environments.

\begin{figure}
\includegraphics[width=80mm,height=140mm ]{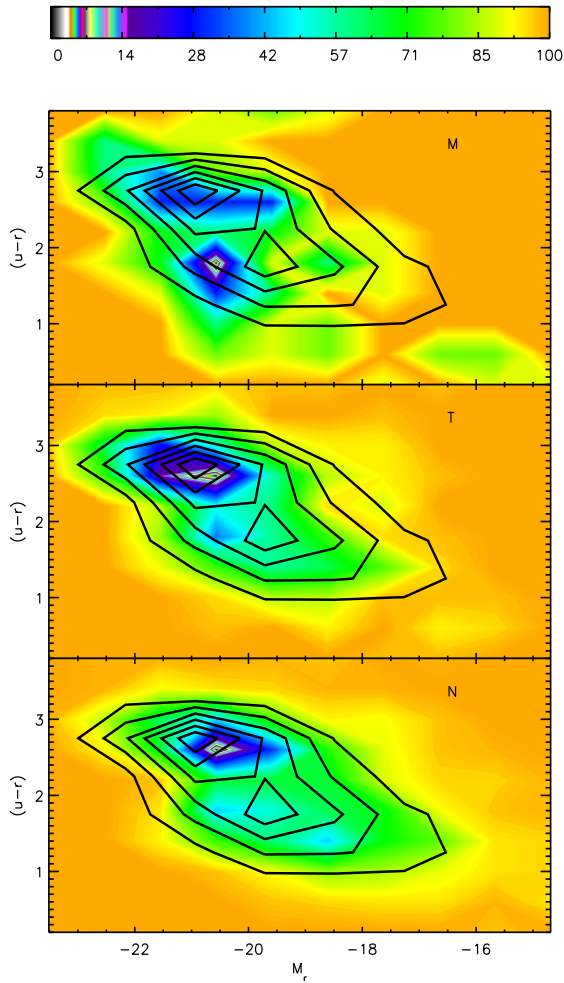}
\caption{Color-magnitude diagrams for merging, tidal and
non disturbed pairs (upper, middle and lower panels, respectively, in a color scale corresponding to
the percentage of objects as shown in the key). 
For comparison, the black solid lines enclose 14$\%$, 28$\%$, 42$\%$, 57$\%$, 71$\%$ and 85$\%$ 
of the isolated galaxies in the control sample.}
\label{col}
\end{figure}

Fig.\ref{col} shows color-magnitude diagrams for different classes of pairs, $M$, $T$ and $N$ (upper, middle and lower panels, respectively) and for galaxies in the control sample (black contours in the three panels; hereafter we show the control sample results in all figures).
As expected, galaxies in the control sample show a bimodal distribution, with a high concentration in the red population.
This behavior is indicating that galaxies in high density environments, such as groups and clusters, are preferentially red objects.
Merging galaxies show a remarkable difference in the color-magnitude relation with respect to galaxies without near companions in the control sample where, for the same magnitude, they show a more scattered distribution, spreading toward blue and red populations.
Tidal and non disturbed pairs show a similar color-magnitude distribution to that of galaxies in the control sample (middle and lower panels, respectively).

\begin{figure}
\includegraphics[width=80mm,height=140mm ]{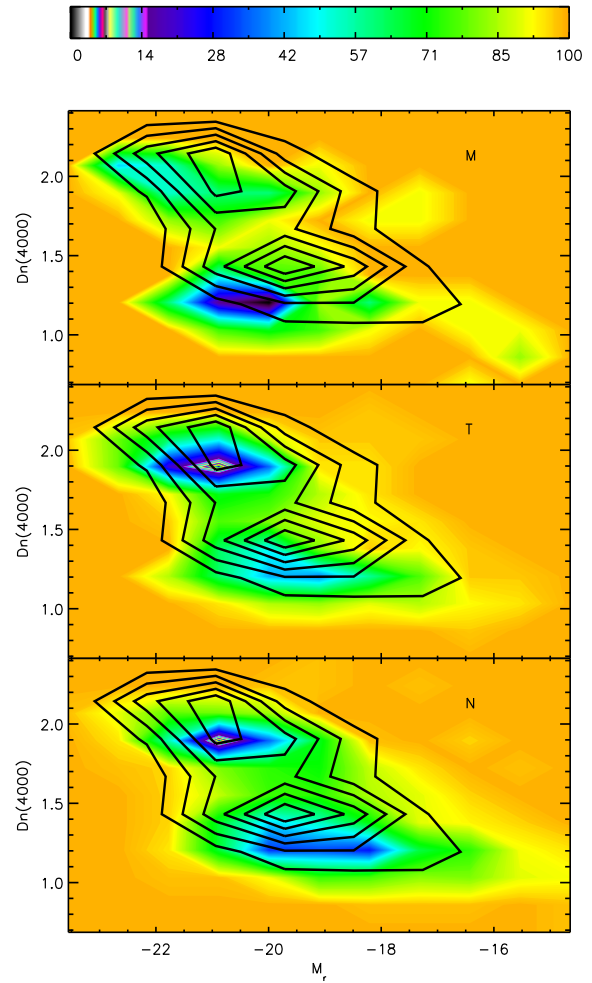}
\caption{Stellar age indicator, $D_n(4000)$ versus $M_r$ for merging, tidal and
non disturbed pairs (upper, middle and lower panels, respectively, in the color scale shown in the key). 
For comparison, the black solid lines enclose 14$\%$, 28$\%$, 42$\%$, 57$\%$, 71$\%$ and 85$\%$ 
of the isolated galaxies in the control sample.}
\label{dnmr}
\end{figure}

To complement the previous analysis, Fig.\ref{dnmr} shows the stellar age indicator, 
$D_n(4000)$, as a function of $M_r$ for different pair classes ($M$, $T$ and $N$; 
upper, middle and lower panels, respectively) and for the control sample
(black solid contours).  
The group members without near neighbors in the control sample show two clear
stellar populations, young and old. 
However, it can be appreciated that galaxy pairs show a clear excess in the young stellar 
population sequence for the same galaxy magnitude, with respect to the control sample.
This tendency is more significant in merging systems.

\begin{figure}
\includegraphics[width=95mm,height=100mm ]{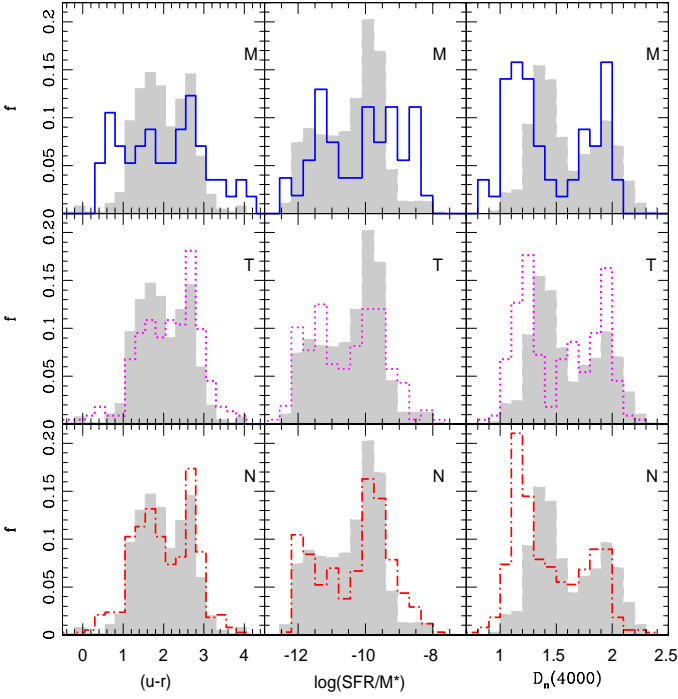}
\caption{ Distributions of $u-r$ (left panels), $log(SFR/M^*)$ (medium panels) and $D_n(4000)$ 
(right panels) for different pair categories (solid lines): $M$, $T$ and $N$ (upper, middle and lower panels, respectively) and the control sample (full surface).
}
\label{histcoldn}
\end{figure}

In Fig.~\ref{histcoldn}, we show the distributions of the $u-r$ color (left panels), 
$log(SFR/M^*)$ (medium panels) and the stellar age indicator $D_n(4000)$ (right panels) for galaxy pairs 
and their isolated counterparts (galaxies in the control sample). 
We find that $T$ and $N$ systems (middle and lower panels, respectively), 
show a more similar color distribution to those of the control sample 
than $M$ galaxy pairs. 
Interestingly, the color distribution of merging galaxy pairs 
and the control sample (upper panel), exhibit significant differences, with
pairs having a larger fraction of red and blue members.
This behavior could suggest that strong galaxy interactions are a powerful mechanism, responsible for the fast and efficient transformation from blue to red colors in dense environments, compared to the less efficient interactions of isolated galaxies in the same density regions.
In addition, the natural interpretation of the blue excess shown in pairs in groups is associated with the star formation activity triggered by galaxy-galaxy interactions (e.g Alonso et al. 2004; Perez et al. 2009).

The red excess in galaxy pairs has been reported by different authors (e.g. Alonso et al. 2006; Perez et al. 2009; Darg et al. 2010) but has no obvious interpretation. 
The red galaxy pairs could be accounted for by dusty, obscured star-forming systems. 
In this case, the slight spread at the red end might be due to increased extinction by 
the extra dust of the perturbing neighbor. 
To test this hypothesis we studied color-color diagrams ((u-r).vs.(i-z)) for pairs ($M$, $T$ and $N$) and isolated galaxies in the control sample. For the analysis we selected edge-on ($b/a<0.4$) and bright galaxies ($M_r<-20.0$), at $z <0.1$, and for both samples we obtained diagrams with no appreciable differences. In a further analysis we divided the samples in ranges of $b/a$ and we compared their color-color diagrams for each range. Unfortunately with the available data it is not possible to confirm if the dust is the responsible for the red and old tails in disturbed pairs.
Another possibility for the red excess is outlined by Patton et al. (2011).  They computed galaxy colors using 
the Galaxy Image 2D (GIM2D, Simard et al. 2002), finding no significant excess ($<$ 1$\%$) of extremely red galaxies in close pairs. The authors suggest that this is due to improved photometry, given that they do find an excess of extremely red close pair members if they replace the GIM2D colors with those from the SDSS database; they also find that the new photometry does not affect the fractions of extremely blue galaxies in pairs.
On the other hand, different authors (e.g. Kauffmann et al. 2004; Baldry et al. 2006; Cooper et al. 2006;
Skibba et al. 2009; Lin et al. 2010; Patton et al. 2011) show that paired galaxies are found in higher density environments than their control sample and that interacting galaxies are redder before the interaction starts, 
due to the older stellar population present in higher density environments. 
However, pairs and isolated galaxies in our samples are selected to the equal density environments: i.e. groups and clusters.
There is another plausible alternative to explain this result.
Red galaxies in pairs within groups and clusters have been stripped of their gas reservoir and show old stellar populations so that the interactions are now unable to trigger further star formation activity. 
However, isolated galaxies in the control sample could include merger remnants
as a consequence of the recently triggered star formation event, reflecting bluer colors,
as is discussed by Perez et al. (2009).

The medium and right panels of Fig.~\ref{histcoldn} show a significantly higher star formation activity and younger stellar population in the three classes of interacting pairs with respect to isolated galaxies in the control sample.
The correlation between colors, star formation rate and stellar population indicates that the physical mechanism responsible for the color transformation also operates by changing the star formation activity and stellar populations in galaxies.

\subsection{Dependence on the group-centric distance and the host group luminosity}

\begin{figure}
\includegraphics[width=70mm,height=90mm ]{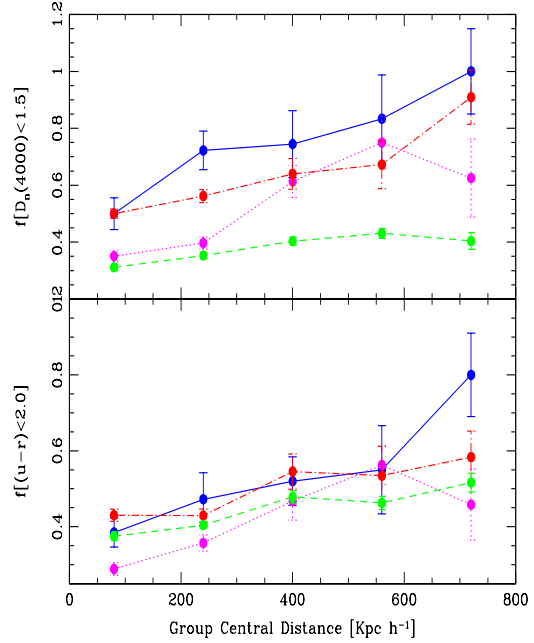}
\caption{Fraction of galaxies with $D_n(4000)<1.5$ (upper panel) and 
$(u-r)<2.0$ as a function of the group-centric distance, for pairs classified 
as $M$, $T$ and $N$ (solid, dotted and dot-dashed lines, respectively) and the control sample (dashed lines).}
\label{DnColdc}
\end{figure}

With the aim to understand the behavior of the colors and stellar populations in galaxy pairs within high density environments, we measured
the blue and young stellar population fractions with respect to the group centric distance.
In Fig.~\ref{DnColdc} we show the fraction of star forming galaxies with $D_n(4000)< 1.5$
 (upper panel) and $u-r < 2.0$ (lower panel) as a function of group-centric distance 
for different classes of pairs and for the control sample.
As can be seen the fraction of star forming galaxies decreases toward the group center.
Similarly, van den Bosch et al. (2008) found that redder satellite galaxies are somewhat more concentrated.  Also, Blanton \& Berlind (2007) showed that generally, the red galaxies are more clustered on small scales ($\approx$ 100-300 kpc $h^{-1}$).
As is well known, the central regions of the groups and clusters of galaxies are mainly populated by spherical objects, without gas, and truncated SFRs.
It can be also appreciated that galaxy pairs show a higher star 
formation activity (low $D_n(4000)$ and blue colors) than galaxies without 
a close companion in the control sample. 
This tendency is more significant in merging pairs.
This fact could indicate that merging pairs in high density regions are still subject to a powerful mechanism that triggers star formation activity.
On the other hand, we can appreciate a higher fraction of $T$ pairs with redder colors and old stellar populations. 
These results clearly show the longer timescale of morphological disturbances with respect to that of tidally induced star formation. 
Thus, these pairs may have old (reddened) stellar populations and still show strong signs of a past interaction.
Also, the strong tidal features can be associated to disruptive effects present in some tidal interactions which would lead to lower gas densities and therefore lower star formation rates in these systems.

\begin{figure}
\includegraphics[width=70mm,height=90mm ]{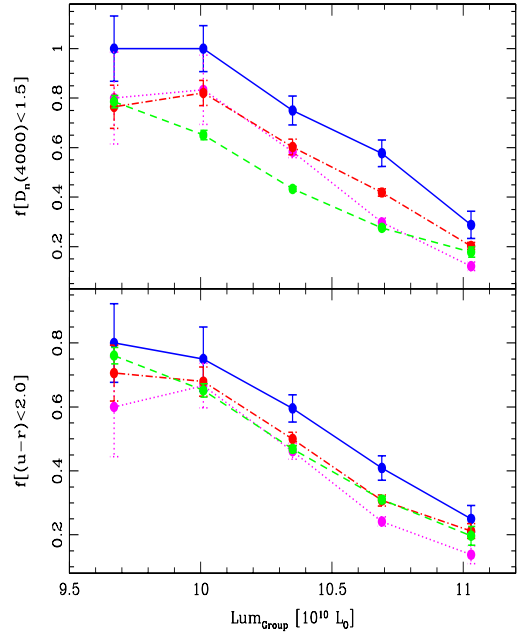}
\caption{Fraction of galaxies with $D_n(4000)<1.5$ (upper panel) and $(u-r)<2.0$ as a 
function of the group total luminosity for pairs classified as $M$, $T$ and $N$ 
(solid, dotted and dot-dashed lines, respectively) and the control sample (dashed lines).}
\label{DnColMr}
\end{figure}

Fig. \ref{DnColMr} shows the fraction of star forming and blue galaxies, 
$D_n(4000)< 1.5$ (upper panel) and $u-r < 2.0$ (lower panel) for pairs as a 
function of the host group luminosity. 
We also show the results for the control sample (dashed lines).
As can be seen, the fraction of star forming galaxies decreases towards high group luminosities.
In addition, we can see that $M$ galaxy pairs have a systematically higher fraction of young stellar
populations and bluer colors than $T$ and $N$ pairs, and galaxies in the control sample.
However, pairs with signs of tidal interactions 
present a lower star formation activity than
galaxies without close companions.
Within the densest regions (more luminosity groups), 
$M$, $T$ and $N$ pairs show similar SF activity than control galaxies, suggesting that in the densest environments the effects of pair interactions become overwhelmed by the intracluster processes.

\section{The effect of the group global environment}

In this section we analyze the effect of the group global environment on galaxy pairs in groups.
For this purpose, we calculate the group global environment parameter, 
$\Sigma_{GR}$=$5/\pi*(d^5_{GR})^2$, 
where $d^5_{GR}$ is the distance to the 5$^{th}$ nearest
group neighbor, which is calculated 
between the centers of groups.
Group neighbors have been chosen to have luminosities above a certain threshold
($Lum_{Group} > 10^{10.6} L_{\sun}$) and with a radial velocity difference lesser 
than 3000 km $s^{-1}$.

\begin{figure}
\includegraphics[width=110mm,height=110mm ]{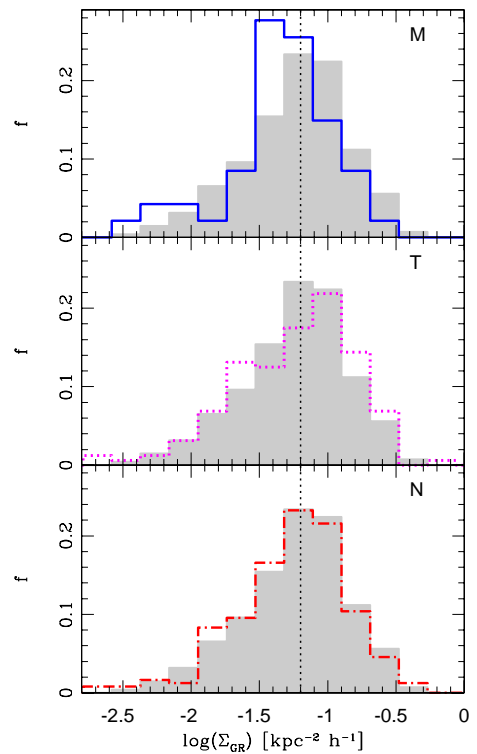}
\caption{Distribution of $log(\Sigma_{GR})$ for different pair categories
$M$, $T$ and $N$ (upper, middle and lower panels, 
respectively) and the control sample (full surface).}
\label{Hsig}
\end{figure}

The influence of the group global environment on the different classes of pairs can be seen 
in Fig.\ref{Hsig} in the histograms of 
$log(\Sigma_{GR})$ for $M$, $T$ and $N$ pairs (upper, medium and lower panels respectively), and galaxies in the control sample (full surface).
We also define the low and high environment classes by selecting two  
ranges of $\Sigma_{GR}$ values in order to have equal number of galaxies in these two classes 
in the control sample. 
This density threshold is log($\Sigma_{GR}$)=-1.2 (represented by the dotted vertical 
line in Fig.\ref{Hsig}).
From this figure it can be appreciated that galaxy pairs 
show an excess of low group global environments  
with respect to the control sample. 
This behavior is more significant in $M$ pairs than in $T$ and $N$ systems ($65 \pm 7 \%$, $48 \pm 6 \%$ and $55 \pm 5 \%$ respectively).

We  explored the stability of the results show in this figure, depending on the 
particular choice of global density estimator. 
We obtained $\Sigma_{GR}$, using $d^5_{GR}$, with two different group luminosity thresholds: $Lum_{Group} > 10^{10.5} L_{\sun}$ and $Lum_{Group} > 10^{10.3} L_{\sun}$. 
We also tested the results $\Sigma_{GR}$ using the distance to the 4$^{th}$ group 
neighbor instead to the 5$th$ group neighbor.  
Taking into account the similarity of the results of these tests within 10$\%$, we conclude that our conclusions are robust against
the particular definition of global group density  environment.

\begin{figure}
\includegraphics[width=70mm,height=90mm ]{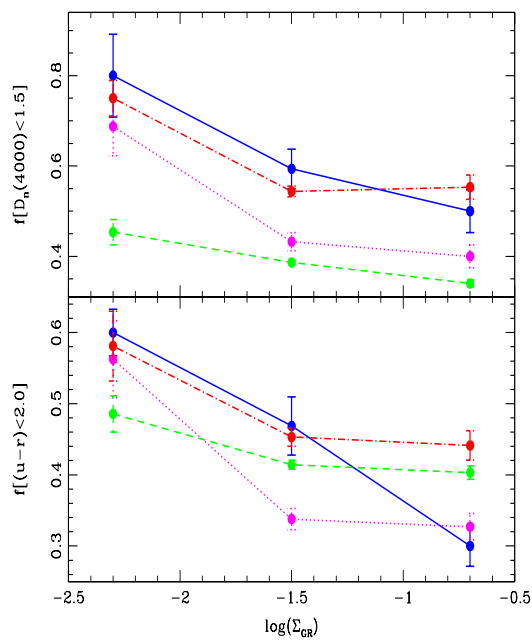}
\caption{Fraction of galaxies with $D_n(4000)<1.5$ (upper panel) and $(u-r)<2.0$ 
(lower panel) as a function of $log(\Sigma_{GR})$, for pair galaxies classified as 
$M$, $T$ and $N$ (solid, dotted and dot-dashed lines, respectively) and the control sample 
(dashed line).}
\label{Dnsgr}
\end{figure}

In order to analyze the behavior of colors and stellar populations in pair systems and in galaxies in the control sample in groups and clusters with respect to the global density environment, we calculated the blue and young stellar population fractions 
($u-r < 2.0$ and $D_n(4000)< 1.5$, respectively)
as a function of $\Sigma_{GR}$.
As can be seen (Fig.~\ref{Dnsgr}), in both samples (pair and control), the fraction of star forming galaxies decreases towards  higher density global environments.
It can also be appreciated that galaxy pairs show a higher star formation activity, more efficient in $M$ systems (low $D_n(4000)$ and bluer colors), than galaxies without a close companion in the control sample. 
These results may indicate that young/old groups and clusters are associated with regions of low/high global overdensity. 
In this context it is interesting to note that galaxy pairs play a fundamental role driving this relationship.
In the same direction, Cooper et al. (2010) using samples drawn from the SDSS, studied the relationship between density and galaxy properties on the red sequence, finding that galaxies with older stellar populations favor regions of higher overdensity relative to galaxies of similar color and luminosity.
In the study by Wang et al. (2007), they find that old, low mass halos at z=0 are associated to higher overdensities. More recently, Lacerna \& Padilla (2011) also find a population of halos that include nearby low mass halos in their original density peak, particularly for old objects.
Our results confirm these previous findings in observational data.

\section{Summary and Conclusions}

We have analyzed the properties of galaxy pairs
in high density environments corresponding to groups and clusters of galaxies.
We summarize the following main conclusions.

(i) We constructed the group galaxy pair catalog and we classified the pair systems in three different categories,
$M$, pairs with evidence of an ongoing merger process, $T$, pairs with signs of tidal interactions, 
and $N$, pairs showing no evidence of distorted morphologies.
By using a Monte Carlo simulation we estimate the contamination by spurious pairs in groups and clusters,
finding that the statistics are clearly dominated by real pairs.
{\it We find that about 9 $\%$ of galaxy pairs are classified as 
$M$, 33 $\%$ as $T$ and 58 $\%$ as $N$.} 
This suggests that mergers are more likely to occur in less-dense groups, where the relative velocities of member galaxies are lower than in more rich groups/clusters 
($\approx$ 1000 km s$^{-1}$; Toomre \& Toomre 1972, Dressler 1980).

(ii) We find that galaxy pairs classified as $M$, $T$ and $N$ are more concentrated
towards the group centers with respect to the other group galaxy members.
We have also analyzed the luminosity galaxy ranking, finding that 
{\it disturbed pairs ($M$ and $T$) have a preference to be the brightest galaxy in groups/clusters 
with respect to non disturbed galaxy pairs and to all group members.}

(iii) We examined the color-magnitude relation for different classes of pairs ($M$, $T$ and $N$) 
and for the control sample, in high density environments.
Galaxies in the control sample show a high red fraction, as expected 
for galaxies within groups and clusters. $N$ pairs show a similar color-magnitude distribution. 
{\it An interesting effect is found for $M$ galaxy pairs, 
where their color distribution exhibits significant differences 
detected in both color tails, with a clear excess of extremely blue and red objects.} 
This tendency could indicate that merging systems within groups and clusters could have 
experienced a more rapid transition from blue to red colors, while isolated galaxies in the control sample and non disturbed pairs undergo a more inefficient transformation.

(iv) Consistent with our analysis on colors, the distribution of star formation rate and 
stellar populations of $M$, $T$ and $N$ galaxy pairs show a clear excess of higher star 
formation activity and young galaxies with respect to
galaxies in the control sample. 
This behavior is more noticeable in merging systems.
{\it The correlation found between bluer colors higher star formation rate and younger 
stellar populations in disturbed pairs 
suggests that strong interactions produce an important effect in modifying galaxy properties 
in dense environments.}

(v) We also studied the fraction of star forming galaxies (blue colors and younger stellar 
population) as a function of the group-centric distance in different pair classes and in the control sample.
We find that the number of star forming galaxies decreases toward the group center as expected, but
{\it galaxy pairs show a higher star formation activity than galaxies without a close companion 
within groups/clusters and that this is more significant in merging pairs.}   
We speculate that this effect could indicate that in high density regions, 
strong galaxy-galaxy interactions are a powerful mechanism that triggers star formation activity.

(vi) We have detected a significant correlation between the fraction of star forming galaxies 
and the host group luminosity,
where {\it the number of star forming objects increases towards low luminosity groups.}
Also, $M$ galaxy pairs have a systematically higher fraction of younger stellar
populations and bluer colors than the other pair categories and galaxies in the control sample.

(vii) We also found that {\it galaxy pairs prefer to be within groups in low density global 
environments}, with respect to galaxies of the corresponding control sample. 
{\it This behavior is more significant in $M$ pairs} ($65 \pm 7 \%$) with respect to $T$ and $N$ systems ($48 \pm 6 \%$ and $55 \pm 5 \%$, respectively).

We have also analyzed the relation between star forming galaxies and the group global environment for pairs and galaxies in the control sample. From this study we conclude that, in particular,
{\it interacting galaxies with a  dominant blue, young stellar population are preferentially in groups within low density global environments.}
Since young groups and clusters are associated with
low global environments we find that this is in agreement with the assembly bias effect 
as this ensures a continuous supply of infalling material, which in turn favors star formation activity.  
In this context the history of galaxy interactions play a fundamental physical role in driving this behavior.

\begin{acknowledgements}

This work was partially supported by the Consejo Nacional de Investigaciones
Cient\'{\i}ficas y T\'ecnicas and the Secretar\'{\i}a de Ciencia y T\'ecnica 
de la Universidad Nacional de San Juan.
Nelson Padilla is supported by Centro de Astrof\'{\i}sica FONDAP, BASAL-CATA and Fondecyt Regular 1110328.

Funding for the SDSS has been provided by the Alfred P. Sloan
Foundation, the Participating Institutions, the National Science Foundation,
the U.S. Department of Energy, the National Aeronautics and Space
Administration, the Japanese Monbukagakusho, the Max Planck Society, and the
Higher Education Funding Council for England. The SDSS Web Site is
http://www.sdss.org/.

The SDSS is managed by the Astrophysical Research Consortium for the
Participating Institutions. The Participating Institutions are the American
Museum of Natural History, Astrophysical Institute Potsdam, University of
Basel, University of Cambridge, Case Western Reserve University,
University of
Chicago, Drexel University, Fermilab, the Institute for Advanced Study, the
Japan Participation Group, Johns Hopkins University, the Joint Institute for
Nuclear Astrophysics, the Kavli Institute for Particle Astrophysics and
Cosmology, the Korean Scientist Group, the Chinese Academy of Sciences
(LAMOST), Los Alamos National Laboratory, the Max-Planck-Institute for
Astronomy (MPIA), the Max-Planck-Institute for Astrophysics (MPA), New Mexico
State University, Ohio State University, University of Pittsburgh, University
of Portsmouth, Princeton University, the United States Naval Observatory, and
the University of Washington.

We thank Laura Balmaceda for her useful advices with IDL codes.

We thank Ramin Skibba for a detail revision that helped to improve this paper.

\end{acknowledgements}

\end{document}